\documentclass{ws-mpla}
\usepackage[super]{cite}
\usepackage{graphicx}

\begin{document}

\markboth{Brian Slovick}
{Renormalization of Einstein gravity through a derivative-dependent field redefinition}

\title{RENORMALIZATION OF EINSTEIN GRAVITY THROUGH A DERIVATIVE-DEPENDENT FIELD REDEFINITION}

\author{\footnotesize BRIAN SLOVICK}

\address{Applied Optics Laboratory, SRI International, Menlo Park, California 94025, United States \\
brian.slovick@sri.com}

\maketitle

\begin{abstract}
This work explores an alternative solution to the problem of renormalizability in Einstein gravity. In the proposed approach, Einstein gravity is transformed into the renormalizable theory of four-derivative gravity by applying a field redefinition containing an infinite number of higher derivatives. It is also shown that the current-current amplitude is invariant with the field redefinition, and thus the unitarity of Einstein gravity is preserved.
\end{abstract}

\ccode{04.60.-m,04.60.Gw,04.50.Kd}

\section{Introduction}

The development of a quantum field theory of gravity based on the Einstein-Hilbert (Einstein) Lagrangian has been problematic because the traditional methods of renormalization cannot be used to eliminate the ultraviolet divergences that appear in perturbation theory \cite{tHooft1974a,Alvarez1989}. Nonrenormalizable terms appear at two loops \cite{Goroff1985} or at one loop when coupled to matter \cite{tHooft1974a}. Alternatively, generalizations of the Einstein Lagrangian that include higher-derivative terms, namely $R_{\mu\nu}R^{\mu\nu}$ and $R^2$, are renormalizable to all orders in perturbation theory \cite{Stelle1977,Voronov1984,Antoniadis1986,Buchbinder1992,Berredo-Peixoto2005}, and the dimensionless couplings of the higher-derivative terms are asymptotically free \cite{Fradkin1982,Tomboulis1980,Avramidi1986}. Also, the essential dimensionless coupling given by the product of the cosmological constant $\Lambda$ and Newton's constant $G$ is claimed to be asymptotically free \cite{Fradkin1982,Codello2006,Julve1978}.

Despite these desirable properties, higher-derivative gravity has a major drawback: in flat-space perturbation theory the higher-derivative terms give rise to a massive spin-two ghost, so the theory is not unitary \cite{Stelle1977,Buchbinder1992,Berredo-Peixoto2005,Tomboulis2015}. It has been suggested that higher-order loop effects may render the massive ghost unstable \cite{Antoniadis1986,Shapiro2015,Modesto2016}, making the theory unitary for asymptotic states, but a rigorous proof of this is lacking.

It is now understood that the Einstein Lagrangian and its higher-derivative extensions may be regarded as the lowest-order terms in the effective field theory of general relativity \cite{Burgess2004}, the theory containing all generally-covariant functions of the metric and its derivatives \cite{Weinberg2009}. One approach for studying the asymptotic behavior of an effective field theory,  referred to as asymptotic safety, is to show that only a subset of the possible couplings are essential, and that they are attracted to a fixed point in the ultraviolet \cite{Weinberg1979,Niedermaier2006,Percacci2009,Christiansen2014,Dona2014}. Non-Gaussian (i.e., non-zero) fixed points have been found by dimensional continuation \cite{Weinberg1979,Kawai1993,Niedermaier2003}, the $1/N$ approximation \cite{Smolin1982,Percacci2006}, the lattice approach \cite{Ambjorn2004,Ambjorn2008}, and various truncations of the functional renormalization group equation \cite{Reuter1998,Lauscher2002,Reuter2002,Niedermaier2009,Benedetti2009}. 

However, by definition an asymptotically safe effective-field theory of gravity will include higher-derivative terms with essential couplings, so the corresponding $S$-matrix may not be unitary in flat-space perturbation theory \cite{Niedermaier2006}. This problem may be avoided if the renormalization group flow near the non-Gaussian fixed point drives the ghost mass to infinity \cite{Julve1978,Benedetti2009}. Another possibility is that the ghost pole in the propagator of the truncated effective Lagrangian is an artifact of the truncation \cite{Weinberg1979,Tomboulis2015}. For example, it has been shown that unitarity arises only when higher-derivative terms of all orders are included \cite{Biswas2012,Shapiro2015,Tomboulis2015}. However, because such actions arise from the expansion of entire functions, they are nonlocal. Until these issues are resolved, it is important to develop alternative methods to study quantum effects in gravity.

This paper explores an alternative method of renormalizing Einstein gravity based on field redefinitions. The equivalence theorem states that physical observables such as amplitudes and $S$-matrix elements are independent of field redefinitions \cite{Chisholm1961,Kallosh1973,Wudka1994,Arzt1995,Einhorn2001,Weinberg2005}. A simple example is a linear field redefinition $\phi \rightarrow Z^{1/2} \phi$, which rescales the propagator by $Z^{-1}$ and the sources by $Z^{1/2}$. Meanwhile, the current-current amplitude, given by the product of two sources and the propagator, is independent of $Z$ \cite{Wudka1994,Arzt1995}. The property of renormalizability, on the other hand, is determined by the derivative or momentum dependence of the propagator and vertices, which in general is not invariant under field redefinitions \cite{tHooft1973b,tHooft1974a,Wudka1994,Arzt1995}. For example, a derivative-dependent field redefinition $\phi \rightarrow Z^{1/2}(k) \phi$ would alter the momentum dependence of the propagator and vertices, while the current-current amplitude, and thus unitarity, would remain invariant. It follows that a derivative-dependent field redefinition can alter renormalizability without affecting unitarity.

Drawing from recent results, \cite{Slovick2013} in this paper this invariance is exploited to obtain a quantum theory of Einstein gravity that is both renormalizable and unitary. Specifically, it is shown that Einstein gravity can be transformed into the renormalizable theory of four-derivative gravity by applying a field redefinition that contains an infinite number of higher derivatives. It is further shown that the current-current amplitude, which embodies the property of unitarity, is invariant with the field redefinition. Thus, the field redefintion renders the theory renormalizable while preserving the unitarity of the Einstein theory.

The following calculations assume natural units, a metric signature of (+ - - -), curvature tensor of $R^\alpha_{\beta \gamma \delta}=-\partial_\delta \Gamma^\alpha_{\beta \gamma}+...$, Ricci tensor defined by $R_{\mu \nu}=R^\alpha_{\mu \nu \alpha}$ and scalar curvature by $R=g^{\mu\nu}R_{\mu\nu}$, where $g_{\mu\nu}$ is the metric tensor. 

\section{Field redefinition and renormalizability}
The classical action of Einstein gravity is
\begin{equation}
\mathcal{S}\equiv \int d^4x \mathcal{L}_g(x)=\frac{2}{\kappa^2}\int d^4x \sqrt{-g} R,
\end{equation}
where $\kappa^2=32\pi G$. 
Consider the following local field redefinition of the metric:
\begin{eqnarray}
g^{\mu\nu}\rightarrow && g^{\mu\nu}+\kappa^2 f_{(1)}^{\mu\nu}, \nonumber\\ 
f_{(1)}^{\mu\nu} = && a R^{\mu\nu}+b Rg^{\mu\nu}.
\end{eqnarray}
The Einstein Lagrangian transforms as
\begin{equation}
\mathcal{L}_g \rightarrow \mathcal{L}'_g = \mathcal{L}_g+\frac{\delta \mathcal{L}_g}{\delta g^{\mu\nu}} \kappa^2 f_{(1)}^{\mu\nu}+\mathcal{O}(\kappa^4),
\end{equation}
where $\mathcal{O}(\kappa^4)$ denotes terms with six or more derivatives. Equation (3) becomes 
\begin{equation}
\mathcal{L}'_g= \sqrt{-g} \left[ \frac{2}{\kappa^2}  R +a R^2_{\mu\nu}-\frac{1}{2} (a+2b) R^2 \right] +\mathcal{O}(\kappa^4).
\end{equation}
The field redefinition has introduced terms with four derivatives of the metric. This leads to propagator and vertex functions, respectively, which vary as $k^{-4}$ and $k^4$ at large momentum. Therefore, to $\mathcal{O}(\kappa^4)$ the transformed Lagrangian is renormalizable in four dimensions. A more detailed proof of the perturbative renormalizability of $\mathcal{L}'_g$ has been provided by Stelle \cite{Stelle1977}.

Because Einstein gravity is nonpolynomial in the metric, there will be an infinite number of terms in the expansion of Eq. (3). The $\mathcal{O}(\kappa^4)$ terms will be cubic and higher in the metric. As a result, the degree of divergence of the vertex functions will be unbounded and the theory will no longer be renormalizable. To maintain renormalizability at higher orders, the field redefinition must be supplemented with additional higher derivative terms as
\begin{equation}
g^{\mu\nu}\rightarrow g^{\mu\nu}+\kappa^2 f_{(1)}^{\mu\nu}+\kappa^4 f_{(2)}^{\mu\nu}+\mathcal{O}(\kappa^6),
\end{equation}
where $f_{(1)}^{\mu\nu}$ is given by Eq. (2). In this case, the Einstein Lagrangian transforms as
\begin{equation}
\mathcal{L}_g \rightarrow \mathcal{L}_g+\frac{\delta \mathcal{L}_g}{\delta g^{\mu\nu}}\left(  \kappa^2 f_{(1)}^{\mu\nu}+\kappa^4 f_{(2)}^{\mu\nu} \right)+\frac{1}{2}\frac{\delta^2 \mathcal{L}_g}{\delta g^{\alpha\beta} \delta g^{\mu\nu}} \kappa^4 f_{(1)}^{\mu\nu} f_{(1)}^{\alpha\beta}+\mathcal{O}(\kappa^6).
\end{equation}
As shown above, terms of order $\kappa^2$ transform Einstein gravity into the renormalizable theory of four-derivative gravity. The role of $f_{(2)}^{\mu\nu}$ is to cancel the higher derivative vertex functions generated by $\kappa^4$ terms such that the transformed Lagrangian remains equivalent to the Lagrangian of four-derivative gravity. This leads to the condition
\begin{equation}
\frac{\delta \mathcal{L}_g}{\delta g^{\mu\nu}} f_{(2)}^{\mu\nu}+\frac{1}{2}\frac{\delta^2 \mathcal{L}_g}{\delta g^{\alpha\beta} \delta g^{\mu\nu}} f_{(1)}^{\mu\nu} f_{(1)}^{\alpha\beta}=0,
\end{equation}
which can be solved for $f_{(2)}^{\mu\nu}$ to obtain
\begin{equation}
f_{(2)}^{\mu\nu}=\frac{1}{2R}g^{\mu\nu} \frac{\delta^2 \mathcal{L}_g}{\delta g^{\alpha\beta} \delta g^{\mu\nu}} f_{(1)}^{\mu\nu} f_{(1)}^{\alpha\beta}.
\end{equation}
Using the expression for the second order variation of the Einstein Lagrangian, \cite{Buchbinder1992,Hamber2009} this can be written as
\begin{eqnarray}
&&f_{(2)}^{\mu\nu}=\frac{1}{2R}g^{\mu\nu} \left( -\frac{1}{8}Rf_{\alpha(1)}^\alpha f_{\beta(1)}^\beta +\frac{1}{4}Rf_{\alpha(1)}^\beta f^\alpha_{\beta(1)} - f^\nu_{\beta(1)} f^\beta_{\alpha(1)} R^\alpha_\nu +\frac{1}{2}f^\alpha_{\alpha(1)} f^\nu_{\beta(1)} R^\beta_\nu  \right. \nonumber\\
&&  \left. -\frac{1}{4} \nabla_\nu f^\alpha_{\beta(1)} \nabla^\nu f^\beta_{\alpha(1)}+\nabla_\nu f^\alpha_{\alpha(1)} \nabla^\nu f^\beta_{\beta(1)} -\frac{1}{2} \nabla_\beta f^\alpha_{\alpha(1)} \nabla^\mu f^\beta_{\mu(1)} +\frac{1}{2}\nabla^\alpha f^\nu_{\beta(1)} \nabla_\nu f^\beta_{\alpha(1)}   \right). \nonumber
\end{eqnarray}
This procedure can be applied at each order of the transformation to ensure the transformed Lagrangian is equivalent to the renormalizable Lagrangian of four derivative gravity. The end result is a renormalizable theory obtained from a field redefinition containing an infinite number of higher derivative terms.

It may also be possible to transform the Einstein theory into another renormalizable model, such as non-local nonpolynomial gravity \cite{Biswas2012,Shapiro2015}, or local polynomial superrenormalizable gravity \cite{Asorey1997,Modesto2016,Modesto2016b,Modesto2017}. This leads to potential ambiguity in calculating the quantum corrections. However, according to the equivalence theorem, physical observables such as $S$-matrix elements and beta functions of essential couplings are invariant under arbitrary local field redefinitions \cite{Chisholm1961,Kallosh1973,Wudka1994,Arzt1995,Einhorn2001,Weinberg2005}. Therefore, in principle, all renormalizable models obtained from the Einstein theory by a local field redefinitions are equally valid. Four-derivative gravity is merely the simplest extension of Einstein gravity sufficient to obtain renormalizability.

\section{Propagator}

To probe the unitary properties of the theory, it is necessary to derive the propagator. This process is greatly simplified using the momentum space projection operators for symmetric rank 2 tensors described in the appendix, which project out the spin-0, spin-1, and spin-2 components of the field  \cite{Stelle1977,Antoniadis1986}. Taking the gravitational field as $g_{\mu\nu}=\eta_{\mu\nu}+\kappa h_{\mu\nu}$, in momentum space the quadratic part of $\mathcal{L}_g$ can be written in terms of the projection operators as \cite{Stelle1977,Julve1978,Fradkin1982}
\begin{eqnarray}
\mathcal{L}_g^{(2)}(k)=\frac{1}{2} h^{\mu\nu} k^2(P_{\mu\nu\rho\sigma}^{(2)}-2P_{\mu\nu\rho\sigma}^{(0-s)})h^{\rho \sigma}.
\end{eqnarray}
In the weak field approximation, the field redefinition in Eq. (2) reduces to
\begin{eqnarray}
h^{\mu\nu}\rightarrow &&h^{\mu\nu}+\kappa^2 f^{\mu\nu}_{(1)}, \nonumber\\
f^{\mu\nu}_{(1)} =&&-\frac{1}{2}k^2 \left[ a P_{\kappa\lambda}^{\mu\nu(2)}+\frac{1}{2}(a+6b) P_{\kappa\lambda}^{\mu\nu(0-s)} \right] h^{\kappa\lambda}.
\end{eqnarray}
The Lagrangian transforms as
\begin{equation}
\mathcal{L}_g^{(2)} \rightarrow \mathcal{L}_g'^{(2)}=\mathcal{L}_g^{(2)}+\frac{\delta \mathcal{L}_g^{(2)}}{\delta h^{\mu\nu}}\kappa^2 f^{\mu\nu}_{(1)}+\mathcal{O}(\kappa^4).
\end{equation}
where $\mathcal{O}(\kappa^4)$ represents vertices with six or more derivatives. Noting the orthogonality properties of the projection operators, namely $P^{(i)}_{\mu\nu\rho\sigma}P^{\mu\nu(j)}_{\kappa\lambda}=\delta^{ij}P^{(i)}_{\kappa\lambda\rho\sigma}$, the transformed Lagrangian simplifies to
\begin{equation}
\mathcal{L}_g'^{(2)}= \frac{1}{2} h^{\mu\nu} k^2\left [-\frac{(k^2-m_2^2)}{m_2^2} P_{\mu\nu\rho\sigma}^{(2)} +\frac{2(k^2-m_0^2)}{m_0^2}P_{\mu\nu\rho\sigma}^{(0-s)} \right] h^{\rho\sigma},
\end{equation}
where $m_2^2=2(a\kappa^2)^{-1}$ and $m_0^2=2[(a+6b)\kappa^2]^{-1}$. The invariance of $\mathcal{L}'^{(2)}_g$ under infinitesimal coordinate transformations of the form
\begin{equation}
x^\mu\rightarrow x^\mu+\kappa \epsilon^\mu(x)
\end{equation}
 leads to a gauge invariance
\begin{equation}
h_{\mu\nu}(x)\rightarrow h_{\mu\nu}(x)-\partial_\mu \epsilon_\nu-\partial_\nu \epsilon_\mu,
\end{equation}
which makes the propagator of $\mathcal{L}_g'^{(2)}$ divergent. This issue is resolved by supplementing $\mathcal{L}'^{(2)}_g$ with a gauge-fixing term as
\begin{equation}
\mathcal{L} = \mathcal{L}'^{(2)}_g+\mathcal{L}_{gf}.
\end{equation}
A particularly useful gauge which leads to a propagator in which all parts vary as $k^{-4}$ at large momentum is the so-called Julve-Tonin gauge \cite{Julve1978,Stelle1977,Antoniadis1986,Accioly2000}
\begin{equation}
\mathcal{L}_{gf}= -\frac{1}{2\xi} h^{\mu\nu} k^2\left [\frac{(k^2-m_2^2)}{m_2^2} P_{\mu\nu\rho\sigma}^{(1)} -\frac{(k^2-2m_2^2)}{m_2^2}P_{\mu\nu\rho\sigma}^{(0-w)} \right] h^{\rho\sigma},
\end{equation}
where $\xi$ is a constant. The total Lagrangian can then be written as \cite{Stelle1977,Antoniadis1986,Accioly2000}
\begin{equation}
\mathcal{L}=\frac{1}{2}  h^{\mu\nu} \mathcal{O}_{\mu\nu\rho\sigma} h^{\rho\sigma},
\end{equation}
where
\begin{eqnarray}
\mathcal{O}_{\mu\nu\rho\sigma}&&=-\frac{k^2}{m_2^2}(k^2-m_2^2)P^{(2)}_{\mu\nu\rho\sigma}+\frac{2k^2}{m_0^2}(k^2-m_0^2)P^{(0-s)}_{\mu\nu\rho\sigma}\nonumber\\
&&-\frac{1}{\xi} \left[\frac{k^2}{m_2^2}(k^2-m_2^2)P^{(1)}_{\mu\nu\rho\sigma}-\frac{k^2}{m_2^2}(k^2-2m_2^2)P^{(0-w)}_{\mu\nu\rho\sigma} \right].
\end{eqnarray}
The propagator, obtained by inverting $\mathcal{O}_{\mu\nu\rho\sigma}$, is then
\begin{eqnarray}
D_{\mu\nu\rho\sigma}&&=\mathcal{O}_{\mu\nu\rho\sigma}^{-1} =-\frac{m_2^2}{k^2(k^2-m_2^2)}P^{(2)}_{\mu\nu\rho\sigma}+\frac{m_0^2}{2k^2(k^2-m_0^2)}P^{(0-s)}_{\mu\nu\rho\sigma}\nonumber\\
&&-\xi \left[ \frac{m_2^2}{k^2(k^2-m_2^2)}P^{(1)}_{\mu\nu\rho\sigma}-\frac{m_2^2}{k^2(k^2-2m_2^2)}P^{(0-w)}_{\mu\nu\rho\sigma} \right].
\end{eqnarray}
It can be seen that all parts of the propagator vary as $k^{-4}$ at large momenta.

\section{Current-current amplitude and unitarity}

The unitarity of the theory can be understood by expanding the propagator into partial fractions. For example, for $\xi=0$
\begin{equation}
D^{\xi=0}_{\mu\nu\rho\sigma}=\frac{P^{(2)}_{\mu\nu\rho\sigma}-\frac{1}{2}P^{(0-s)}_{\mu\nu\rho\sigma}}{k^2}-\frac{P^{(2)}_{\mu\nu\rho\sigma}}{k^2-m_2^2}+\frac{\frac{1}{2}P^{(0-s)}_{\mu\nu\rho\sigma}}{k^2-m_0^2}.
\end{equation}
The field redefinition has formally introduced additional massive graviton states. The first term corresponds to the massless spin-2 graviton, while the second and third terms, respectively, correspond to massive spin-2 and spin-0 states. Note that for $m_2, m_0\rightarrow \infty$, $D^{\xi=0}_{\mu\nu\rho\sigma}$ reduces to the propagator of the Einstein theory. The conditions for unitarity at tree level can be determined from the current-current transition amplitude given by \cite{Nunes1993,Pinheiro1996,Accioly2002,Hassan2012}
\begin{equation}
\mathcal{M}(k) = \frac{1}{2} \kappa^2 T^{\mu\nu}(-k) D^{\xi=0}_{\mu\nu\rho\sigma}(k)T^{\rho\sigma}(k),
\end{equation}
where $T_{\mu\nu}$ is the stress-energy tensor. Unitarity requires the imaginary part of the residue of $\mathcal{M}(k)$ at the poles to be positive \cite{Nunes1993,Pinheiro1996,Accioly2002,Hassan2012}. While the residues of the massless spin-2 and the spin-0 state are positive, the residue of the massive spin-2 state is negative, which would normally violate the unitarity condition. However, noting that the sources couple linearly to the fields as $h^{\mu\nu}T_{\mu\nu}$, the linear field redefinition in Eq. (11) also requires the sources in $\mathcal{M}(k)$ to be redefined as  
\begin{equation}
T^{\mu\nu}\rightarrow  T'^{\mu\nu}= T^{\mu\nu}-\frac{1}{2}k^2 \left( \frac{1}{m_2^2} P_{\kappa\lambda}^{\mu\nu(2)}+\frac{1}{m_0^2} P_{\kappa\lambda}^{\mu\nu(0-s)} \right) T^{\kappa\lambda}.
\end{equation}
As a result, the amplitude is invariant under the field redefinition,
\begin{eqnarray}
\mathcal{M}'(k)&& =\frac{1}{2} \kappa^2 T'^{\mu\nu}(-k) D^{\xi=0}_{\mu\nu\rho\sigma}(k)T'^{\rho\sigma}(k) \nonumber\\
&&=\frac{1}{2} \kappa^2 T^{\mu\nu}(-k) \frac{P^{(2)}_{\mu\nu\rho\sigma}-\frac{1}{2}P^{(0-s)}_{\mu\nu\rho\sigma}}{k^2}T^{\rho\sigma}(k).
\end{eqnarray}
That is, only the propagator of the massless spin-2 state appears in the amplitude. Since the imaginary part of the on-shell residue of this portion of the propagator is positive, the unitarity condition is satisfied at tree level.

Beyond tree level, unitarity is preserved provided that the field redefinition is modified to include radiative corrections to the masses. For example, at one-loop order radiative corrections lead to a dressed propagator of the form \cite{Fradkin1982,Modesto2016}
\begin{equation}
D^{\xi=0}_{\mu\nu\rho\sigma}=\frac{P^{(2)}_{\mu\nu\rho\sigma}}{-\frac{k^2}{M_2^2}(k^2-M_2^2)+\alpha_2\kappa^2 k^4 \log{\frac{k^2}{M_2^2}}}+\frac{P^{(0-s)}_{\mu\nu\rho\sigma}}{\frac{2k^2}{M_0^2}(k^2-M_0^2)+\alpha_0\kappa^2 k^4 \log{\frac{k^2}{M_0^2}}},
\end{equation}
where $M_2$ and $M_0$ are the renormalized masses. This dressed propagator is obtained from the field redefinition in Eq. (11) by replacing the bare masses as
\begin{equation}
\frac{1}{m_2^2}\rightarrow \frac{1}{M_2^2} \left( 1-\alpha_2 \kappa^2 M_2^2 \log{\frac{k^2}{M_2^2}} \right), \quad \frac{1}{m_0^2}\rightarrow \frac{1}{M_0^2} \left( 1+\frac{1}{2}\alpha_0 \kappa^2 M_0^2 \log{\frac{k^2}{M_0^2}} \right).
\end{equation}
Importantly, as long as this replacement is also made in the source redefinition of Eq. (22), the contribution of the massive states to the amplitude in Eq. (23) vanishes, and unitarity is preserved at one-loop order.

In addition to the transformation of the Lagrangian and the redefinition of the source, there is a Jacobian associated with the field redefinition. For local transformations, the Jacobian can be written as a ghost Lagrangian of the form \cite{tHooft1973b,Arzt1995}
\begin{equation}
-\mathcal{L}_{ghost}=c \bar{c}+c \frac{\delta f^{\mu\nu}_{(1)}}{\delta h^{\mu\nu}}\bar{c},
\end{equation}
where $c$ and $\bar{c}$ are the ghost fields. Since $f^{\mu\nu}_{(1)}$ is linear in the second derivatives of $h^{\mu\nu}$, the ghost acquires a kinetic term but does not couple to the physical field $h^{\mu\nu}$. Therefore, the ghost contributes only an overall contant to the generating functional and thus has no physical effect.

\section{Summary}

This work aims to develop a quantum theory of gravity that is both unitary and power-counting renormalizable. The approach is to transform Einstein gravity into the renormalizable theory of four-derivative gravity through a field redefinition containing an infinite number of higher derivatives. Importantly, it is also shown that the current-current amplitude is invariant with the field redefinition, and thus the unitarity of the Einstein theory is preserved.

\appendix
\section{Projection operators}
The derivation of the graviton propagator is considerably simplified using the momentum space projection operators for symmetric rank 2 tensors. The complete set of projection operators in momentum space is \cite{Stelle1977,Antoniadis1986}
$$
P^{(2)}_{\mu\nu\rho\sigma}=\frac{1}{2}(\theta_{\mu\rho}\theta_{\nu\sigma}+\theta_{\mu\sigma}\theta_{\nu\rho})-\frac{1}{3}\theta_{\mu\nu}\theta_{\rho\sigma}
$$
$$
P^{(1)}_{\mu\nu\rho\sigma}=\frac{1}{2}(\theta_{\mu\rho}\omega_{\nu\sigma}+\theta_{\mu\sigma}\omega_{\nu\rho}+\theta_{\nu\rho}\omega_{\mu\sigma}+\theta_{\nu\sigma}\omega_{\mu\rho})
$$
$$
P^{(0-s)}_{\mu\nu\rho\sigma}=\frac{1}{3}\theta_{\mu\nu}\theta_{\rho\sigma}
$$
$$
P^{(0-w)}_{\mu\nu\rho\sigma}=\omega_{\mu\nu}\omega_{\rho\sigma}
$$
$$
P^{(0-sw)}_{\mu\nu\rho\sigma}=\frac{1}{\sqrt{3}}\theta_{\mu\nu}\omega_{\rho\sigma}
$$
$$
P^{(0-ws)}_{\mu\nu\rho\sigma}=\frac{1}{\sqrt{3}}\omega_{\mu\nu}\theta_{\rho\sigma}
$$
where $\theta_{\mu\nu}$ and $\omega_{\mu\nu}$, respectively, are the transverse and longitudinal vector projection operators given by
$$
\theta_{\mu\nu}\equiv \eta_{\mu\nu}-k_\mu k_\nu/k^2
$$
$$
\omega_{\mu\nu}\equiv k_\mu k_\nu/k^2
$$
The orthogonality relations are
$$
P^{i-a}P^{j-b}=\delta^{ij}\delta^{ab}P^{j-b}
$$
$$
P^{i-ab}P^{j-cd}=\delta^{ij}\delta^{bc}P^{j-a}
$$
$$
P^{i-a}P^{j-bc}=\delta^{ij}\delta^{ab}P^{j-ac}
$$
$$
P^{i-ab}P^{j-c}=\delta^{ij}\delta^{bc}P^{j-ac}
$$
where $i,j=0,1,2$ and $a,b=s,w$.\\

\providecommand{\noopsort}[1]{}\providecommand{\singleletter}[1]{#1}%


\begin{thebibliography}{10}

\bibitem{tHooft1974a}
G.~'t~Hooft and M.~Veltman.
\newblock One loop divergencies in the theory of gravitation.
\newblock {\em Ann. Inst. Henri Poincar\'{e}}, 20:69--94, 1974.

\bibitem{Alvarez1989}
E.~Alvarez.
\newblock Quantum gravity: an introduction to some recent results.
\newblock {\em Rev. Mod. Phys.}, 61:561--604, 1989.

\bibitem{Goroff1985}
M.~H. Goroff and A.~Sagnotti.
\newblock Quantum gravity at two loops.
\newblock {\em Phys. Lett. B}, 160:81, 1985.

\bibitem{Stelle1977}
K.~S. Stelle.
\newblock Renormalization of higher-derivative quantum gravity.
\newblock {\em Phys. Rev. D}, 16:953--969, 1977.

\bibitem{Voronov1984}
B.~L. Voronov and I.~V. Tyutin.
\newblock On renormalization of r2 gravitation.
\newblock {\em Yad. Fiz.(Sov. Journ. Nucl. Phys.)}, 39:998, 1984.

\bibitem{Antoniadis1986}
I.~Antoniadis and E.~T. Tomboulis.
\newblock Gauge invariance and unitarity in higher-derivative quantum gravity.
\newblock {\em Phys. Rev. D}, 33:2756, 1986.

\bibitem{Buchbinder1992}
I.~L. Buchbinder, S.~D. Odintsov, and I.~L. Shapiro.
\newblock {\em Effective Action in Quantum Gravity}.
\newblock IOP Publishing, 1992.

\bibitem{Berredo-Peixoto2005}
G.~de~Berredo-Peixoto and I.~L. Shapiro.
\newblock Higher derivative quantum gravity with gauss-bonnet term.
\newblock {\em Phys. Rev. D}, 71:064005-- 064020, 2005.

\bibitem{Fradkin1982}
E.~S. Fradkin and A.~A. Tseytlin.
\newblock Renormalizable asymptotically free quantum theory of gravity.
\newblock {\em Nucl. Phys. B}, 201:469--491, 1982.

\bibitem{Tomboulis1980}
E.~Tomboulis.
\newblock Renormalizability and asymptotic freedom in quantum gravity.
\newblock {\em Phys. Lett. B}, 97:77--80, 1980.

\bibitem{Avramidi1986}
I.~G. Avramidi.
\newblock Asymptotic behavior of the quantum theory of gravity with higher
  order derivatives.
\newblock {\em Yad. Fiz}, 44:255--263, 1986.

\bibitem{Codello2006}
A.~Codello and R.~Percacci.
\newblock Fixed points of higher-derivative gravity.
\newblock {\em Phys. Rev. Lett.}, 97:221301--221304, 2006.

\bibitem{Julve1978}
J.~Julve and M.~Tonin.
\newblock Quantum gravity with higher derivative terms.
\newblock {\em II Nuovo Cimento B Series}, 46:137--152, 1978.

\bibitem{Tomboulis2015}
E.~T. Tomboulis.
\newblock Renormalization and unitarity in higher derivative and nonlocal
  gravity theories.
\newblock {\em Mod. Phys. Lett. A}, 30:1540005, 2015.

\bibitem{Shapiro2015}
I.~L. Shapiro.
\newblock Counting ghosts in the ''ghost-free'' non-local gravity.
\newblock {\em Phys. Lett. B}, 744:67, 2015.

\bibitem{Modesto2016}
L.~Modesto.
\newblock Super-renormalizable or finite lee–wick quantum gravity.
\newblock {\em Nucl. Phys. B}, 909:584, 2016.

\bibitem{Burgess2004}
C.~P. Burgess.
\newblock Quantum gravity in everyday life: General relativity as an effective
  field theory.
\newblock {\em Living Rev. Relativity}, 7:5--56, 2004.

\bibitem{Weinberg2009}
S.~Weinberg.
\newblock Effective field theory, past and future.
\newblock {\em arXiv preprint}, arXiv:0908.1964:1--24, 2009.

\bibitem{Weinberg1979}
S.~Weinberg.
\newblock Ultraviolet divergences in quantum theories of gravitation.
\newblock In S.W. Hawking and W.~Israel, editors, {\em General Relativity: An
  Einstein Centenary Survey}. Cambridge University Press, Oxford, 1979.

\bibitem{Niedermaier2006}
M.~Niedermaier and M.~Reuter.
\newblock The asymptotic safety scenario in quantum gravity.
\newblock {\em Living Rev. Relativity}, 9:5--173, 2006.

\bibitem{Percacci2009}
R.~Percacci.
\newblock Asymptotic safety.
\newblock In D.~Oriti, editor, {\em Approaches to quantum gravity}. Cambridge
  University Press, Cambridge, 2009.

\bibitem{Christiansen2014}
N.~Christiansen, D.~F. Litim, J.~M. Pawlowski, and A.~Rodigast.
\newblock Fixed points and infrared completion of quantum gravity.
\newblock {\em Phys. Lett. B}, 728:114, 2014.

\bibitem{Dona2014}
P.~Dona, A.~Eichhorn, and R.~Percacci.
\newblock Matter matters in asymptotically safe quantum gravity.
\newblock {\em Phys. Rev. D}, 89:084035, 2014.

\bibitem{Kawai1993}
H.~Kawai, Y.~Kitazawa, and M.~Ninomiya.
\newblock Ultraviolet stable fixed point and scaling relations in
  (2+$\epsilon$)-dimensional quantum gravity.
\newblock {\em Nucl. Phys. B}, 404:684--714, 1993.

\bibitem{Niedermaier2003}
M.~Niedermaier.
\newblock Dimensionally reduced gravity theories are asymptotically safe.
\newblock {\em Nucl. Phys. B}, 673:131--169, 2003.

\bibitem{Smolin1982}
L.~Smolin.
\newblock A fixed point for quantum gravity.
\newblock {\em Nucl. Phys. B}, 208:439--466, 1982.

\bibitem{Percacci2006}
R.~Percacci.
\newblock Further evidence for a gravitational fixed point.
\newblock {\em Phys. Rev. D}, 73:041501--041504, 2006.

\bibitem{Ambjorn2004}
J.~Ambj{\o}rn, J.~Jurkiewicz, and R.~Loll.
\newblock Emergence of a 4d world from causal quantum gravity.
\newblock {\em Phys. Rev. Lett.}, 93:131301--131304, 2004.

\bibitem{Ambjorn2008}
J.~Ambj{\o}rn, A.~G{\"{o}}rlich, J.~Jurkiewicz, and R.~Loll.
\newblock Nonperturbative quantum de sitter universe.
\newblock {\em Phys. Rev. D}, 78:063544--063560, 2008.

\bibitem{Reuter1998}
M.~Reuter.
\newblock Nonperturbative evolution equation for quantum gravity.
\newblock {\em Phys. Rev. D}, 57:971--985, 1998.

\bibitem{Lauscher2002}
O.~Lauscher and M.~Reuter.
\newblock Flow equation of quantum einstein gravity in a higher-derivative
  truncation.
\newblock {\em Phys. Rev. D}, 66:025026-- 025075, 2002.

\bibitem{Reuter2002}
M.~Reuter and F.~Saueressig.
\newblock Renormalization group flow of quantum gravity in the einstein-hilbert
  truncation.
\newblock {\em Phys. Rev. D}, 65:065016--065041, 2002.

\bibitem{Niedermaier2009}
M.~R. Niedermaier.
\newblock Gravitational fixed points from perturbation theory.
\newblock {\em Phys. Rev. Lett.}, 103:101303--101306, 2009.

\bibitem{Benedetti2009}
D.~Benedetti, P.~F. Machado, and F.~Saueressig.
\newblock Asymptotic safety in higher-derivative gravity.
\newblock {\em Mod. Phys. Lett. A}, 24:2233--2241, 2009.

\bibitem{Biswas2012}
T.~Biswas, E.~Gerwick, T.~Koivisto, and A.~Mazumdar.
\newblock Towards singularity-and ghost-free theories of gravity.
\newblock {\em Phys. Rev. Lett.}, 108:031101, 2012.

\bibitem{Chisholm1961}
J.~S.~R. Chisholm.
\newblock Change of variables in quantum field theories.
\newblock {\em Nucl. Phys.}, 26:469--479, 1961.

\bibitem{Kallosh1973}
R.~E. Kallosh and I.~V. Tyutin.
\newblock The equivalence theorem and gauge invariance in renormalizable
  theories.
\newblock {\em Yad. Fiz.}, 17:190--209, 1973.

\bibitem{Wudka1994}
J.~Wudka.
\newblock Electroweak effective lagrangians.
\newblock {\em Int. J. Mod Phys B}, 9:2301--2361, 1994.

\bibitem{Arzt1995}
C.~Arzt.
\newblock Reduced effective lagrangians.
\newblock {\em Phys. Lett. B}, 342:189--195, 1995.

\bibitem{Einhorn2001}
M.~B. Einhorn and J.~Wudka.
\newblock Effective beta-functions for effective field theory.
\newblock {\em J. High Energ. Phys.}, 2001:1--13, 2001.

\bibitem{Weinberg2005}
S.~Weinberg.
\newblock {\em The Quantum Theory of Fields, Volume I: Foundations}.
\newblock Cambridge University Press, 2005.

\bibitem{tHooft1973b}
G.~'t~Hooft and M.~Veltman.
\newblock Diagrammar.
\newblock Technical Report 73-9, CERN, 1973.

\bibitem{Slovick2013}
B.~A. Slovick.
\newblock Renormalization and asymptotic freedom in quantum gravity through the
  equivalence theorem.
\newblock {\em arXiv preprint}, arXiv:1309.5945, 2013.

\bibitem{Hamber2009}
H.~W. Hamber.
\newblock {\em Quantum Gravitation: The Feynman Path Integral Approach}.
\newblock Springer, 2009.

\bibitem{Asorey1997}
M.~Asorey, J.~L. Lopez, and I.~L. Shapiro.
\newblock Some remarks on high derivative quantum gravity.
\newblock {\em Int. J. Mod. Phys. A}, 12:5711, 1997.

\bibitem{Modesto2016b}
L.~Modesto and I.~L. Shapiro.
\newblock Superrenormalizable quantum gravity with complex ghosts.
\newblock {\em Phys. Lett. B}, 755:279, 2016.

\bibitem{Modesto2017}
L.~Modesto, L.~Rachwal, and I.~L. Shapiro.
\newblock Renormalization group in super-renormalizable quantum gravity.
\newblock {\em arXiv preprint}, arXiv:1704.03988:1--16, 2017.

\bibitem{Accioly2000}
A.~Accioly, S.~Ragusa, H.~Mukai, and E.~de~Rey~Neto.
\newblock Algorithm for computing the propagator for higher derivative gravity
  theories.
\newblock {\em Int. J. Theo. Phys.}, 39:1599, 2000.

\bibitem{Nunes1993}
F.~C.~P. Nunes and G.~O. Pires.
\newblock Extending the barnes-rivers operators to d= 3 topological gravity.
\newblock {\em Phys. Lett. B}, 301:339, 1993.

\bibitem{Pinheiro1996}
C.~Pinheiro, G.~O. Pires, and N.~Tomimura.
\newblock Some quantum aspects of three-dimensional einstein-chern-simons-proca
  massive gravity.
\newblock {\em II Nuovo Cimento B}, 111:1023, 1996.

\bibitem{Accioly2002}
A.~Accioly, A.~Azeredo, and H.~Mukai.
\newblock Propagator, tree-level unitarity and effective nonrelativistic
  potential for higher-derivative gravity theories in d dimensions.
\newblock {\em J. Math. Phys.}, 43:473, 2002.

\bibitem{Hassan2012}
S.~F. Hassan, R.~A. Rosen, and A.~Schmidt-May.
\newblock Ghost-free massive gravity with a general reference metric.
\newblock {\em J. High Energy Phys.}, 2:1, 2012.

\end{thebibliography}
\end{document}